\documentclass[transmag,letterpaper]{IEEEtran}
\newif\ifpeerreview

\peerreviewfalse

\newcommand{\paperID}{32}

\usepackage[switch]{lineno}

\usepackage{cite}
\usepackage{url}
\usepackage{lipsum}

\usepackage{graphicx}
\usepackage{subcaption}

\ifCLASSINFOpdf
\else
\fi

\begin{document}

\ifpeerreview
  \linenumbers
  \linenumbersep 5pt\relax
\fi 

%
\title{Light-Field for RF}


\ifpeerreview
\author{Anonymous Authors}
\else
\author{\IEEEauthorblockN{Manikanta Kotaru\IEEEauthorrefmark{1},
Guy Satat\IEEEauthorrefmark{2},
Ramesh Raskar\IEEEauthorrefmark{2}, and
Sachin Katti\IEEEauthorrefmark{1}}
\IEEEauthorblockA{\IEEEauthorrefmark{1}Stanford University}
\IEEEauthorblockA{\IEEEauthorrefmark{2}Massachusetts Institute of Technology}
}
\fi

\ifpeerreview
\markboth{Anonymous ICCP 2019 submission ID \paperID}%
{}
\else
\fi

\IEEEtitleabstractindextext{%
\begin{abstract}
Most computer vision systems and computational photography systems are visible light based which is a small fraction of the electromagnetic (EM) spectrum. In recent years radio frequency (RF) hardware has become more widely available, for example, many cars are equipped with a RADAR, and almost every home has a WiFi device.
In the context of imaging, RF spectrum holds many advantages compared to visible light systems. In particular, in this regime, EM energy effectively interacts in different ways with matter. This property allows for many novel applications such as privacy preserving computer vision and imaging through absorbing and scattering materials in visible light such as walls.
Here, we expand many of the concepts in computational photography in visible light to RF cameras.
The main limitation of imaging with RF is the large wavelength that limits the imaging resolution when compared to visible light. However, the output of RF cameras is usually processed by computer vision and perception algorithms which would benefit from multi-modal sensing of the environment, and from sensing in situations in which visible light systems fail.
To bridge the gap between computational photography and RF imaging, we expand the concept of light-field to RF. This work paves the way to novel computational sensing systems with RF.

\end{abstract}

\ifpeerreview
\else
\fi
}

\maketitle

\IEEEdisplaynontitleabstractindextext

\section{Introduction}
\IEEEPARstart{T}{raditional} computer vision and computational photography systems are primarily visible light based. This biased focus can be attributed to the sensitivity of our eyes to electromagnetic (EM) energy in the visible spectra, the availability of high quality lenses and dense array of detectors (pixels), and the relatively short wavelength of visible light which results in high resolution images. In parallel to the development of visible light computer vision systems, radio frequency (RF) imaging systems were developed. RF systems are mainly used for communication purposes, and in the form of RADAR for localization, tracking, and velocity estimation. In recent years RF systems have become widely used  e.g. in cars and homes, which motivates the development of imaging solutions in RF. Here, we expand many of the concepts developed for visible light to RF spectrum towards RF cameras.

Imaging with RF holds many applications that challenge computer vision with visible light. Privacy is a significant challenge for many computer vision systems, but it is easily solved with RF since no identifying or exposing information are captured. Visible light is absorbed by many materials that are transparent in RF, thus imaging through occlusions such as walls is easier in RF. Finally, many computer vision systems are safety critical and require redundancy in the form of sensing hardware and perception algorithms. Such redundancy can be achieved by fusing visible light cameras and RF cameras.

Visible light and RF share many properties as they are both EM waves. However, due to the significant difference in wavelength, there are many practical differences for the purposes of imaging. These differences can be considered as a limitation or an advantage depending on the application:

\begin{itemize}
\item Interaction with matter - for example, RF is mostly sensitive to bulk properties of objects, while visible light is sensitive to finer features. This makes it harder to distinguish among different materials with RF. On the other hand, RF propagates through materials that absorb in visible light such as walls.

\item In RF we can easily measure the phase of the wavefront. This is essential since we don't have focusing elements in RF like lenses. Thus, only bare sensor imaging is possible which is fundamentally easier with phase measurement.

\item With RF we can easily measure depth and speed of objects.

\item Due to the larger wavelength of RF, pixels (detectors) are relatively large which limits our ability to construct dense and high resolution sensor arrays.

\end{itemize}

\begin{figure}
  \centering
    \includegraphics[width=1\linewidth]{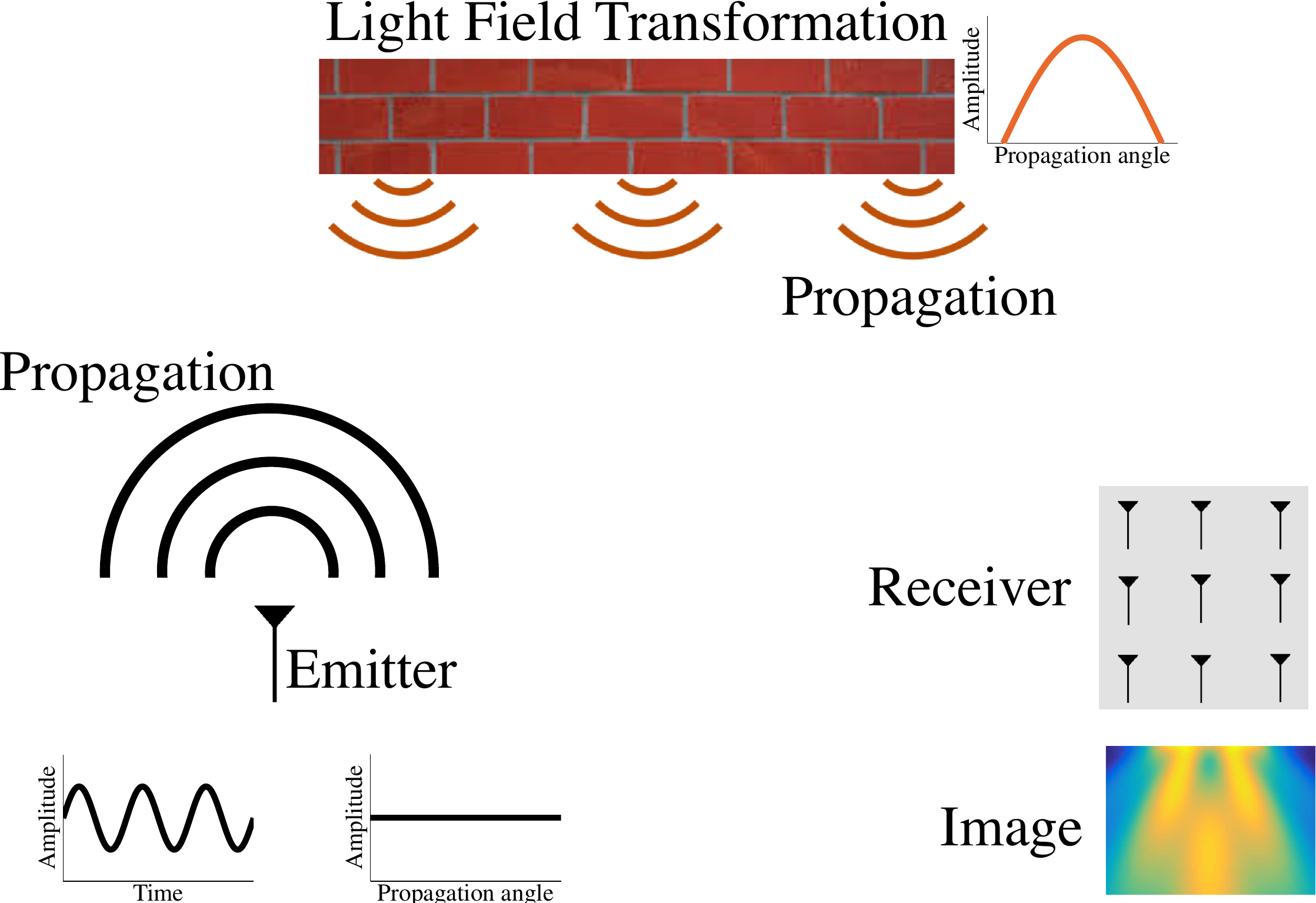}
      \caption{Light-field for RF cameras. We propose a Plenoptic function formulation for RF including  emission,propagation, interaction with planar surfaces, and a rendering technique for RF systems.}
      \label{fig:system}
\end{figure}

\begin{table*}[ht]
\begin{center}
    \caption{Comparison of visible light and RF sensors}
\label{compare}

\small{
    \begin{tabular}{ | p{3.8cm} | p{6.0cm} |  p{6.8cm} |}
    \hline
    & \textbf{Visible light} & \textbf{RF } \\ \hline
    Wavelength & 100s of nms & few cms \\ \hline
    \textbf{Camera} &  & \\ \hline
    Plenoptic function & $L(x,y,z, \xi, \psi, t, f)$, real-valued intensity  & $L(x,y,z, \xi, \psi, t, f)$, complex-valued  amplitude  \\ \hline 
    
    Image & projection of plenoptic function & $L(x_k, y_k, z_k, f)$ \\ \hline

    Sensors & 	 Photodetectors  & Antennas \\ \hline
    
    Field of view & Small & Large; typically omni-directional in azimuth angle \\ \hline
    
    Number of sensors & Few millions & 3-4 in conventional devices \\ \hline
    
    Lens & 	 Yes & No \\ \hline
    
    \textbf{Emitter} &  & \\ \hline
    Coherence & Typically incoherent sources; Lasers are coherent & Coherent sources \\ \hline

     Generated signal & $A x(t)$; $A$ and $x(t)$ are real-valued; $A$ is intensity & $ A \mathrm{e}^{i 2 \pi f_c t } x(t)$, $A$ and $x(t)$ are complex-valued; $A$ is complex amplitude \\ \hline
        
    Carrier frequency  & few 100s of THz & few GHz \\ \hline

    \textbf{Light Field transformation} &  & \\ \hline
    
    Free space propagation & $\frac{A}{r^2} x(t - \frac{r}{c})$  &  $\frac{A}{r} \mathrm{e}^{ i2 \pi f t} \mathrm{e}^{- i2 \pi \frac{r}{\lambda}}$ \\ \hline
    
    BRDF of small light patches & complex function of incoming and outgoing propagation angles; real-valued  &  $\alpha S \cos(\theta) $; complex-valued \\ \hline
    

     \textbf{Receiver} &  & \\ \hline
    
    Image &  $ \int_T  L[x_k, y_k, z_k, \xi, \psi, t, f]$   & $ \int_T \big( \int_{\theta,\psi} L[x_k, y_k, z_k, \xi, \psi, t, f]  \big) \mathrm{e}^{-i2\pi f t - \phi}$ \\ \hline
    
    Carrier phase & Only phase difference is measurable with coherent sources, e.g., in OCT  & Measured \\ \hline
    
    Exposure time & few milliseconds & microseconds \\ \hline

    \end{tabular}
}
    
\end{center}
\end{table*}



    
    
   
    
    

    

Many imaging applications using RF have been demonstrated, however they are application specific (e.g.~\cite{charvat2015time, adib2015capturing}). Here we take a broader approach and discuss, from first principles, the mathematical similarities and concepts describing RF and visible light cameras. We summarize these concepts in Table~\ref{compare}.

This paper highlights the differences and trade-offs between RF and visible light when considering imaging with RF systems. We build upon many of the advances in the computer vision and computational photography communities and expand many of the concepts developed primarily for visible light into RF cameras. The main contributions of our work are:

\begin{itemize}
\item A RF plenoptic function light-field formulation.
\item A light-field transformation for RF that explores the complete path including emission, propagation, interaction with planar surfaces, and array based sensing.

\item A rendering solution for RF systems that accounts for wave effects, without the complexity of specialized simulators. 
\end{itemize}

We first define the notions of RF plenoptic function and RF camera in Sec.~\ref{sec:lightField} and Sec.~\ref{sec:sensor} respectively. As shown in Fig.~\ref{fig:system}, any RF imaging system consists of five components: 1) emission of RF light-field, 2) propagation through free space, 3) transformation of RF light-field as it interacts with objects in the environment, 4) propagation to a receiver, and 5) reception and processing of RF light-field resulting in the RF image. These phenomena are described in Sec.~\ref{sec:source}-Sec.~\ref{sec:receiver}. We conclude with an analysis of the resolution of images that can be obtained from an RF camera in Sec.~\ref{sec:resolution}.

\section{Related Works}\label{sec:relatedworks}

\subsection{Cameras Beyond Visible Light}
Majority of imaging modalities are in the visible part of the EM spectrum. Additional imaging modalities that are not EM based include e.g. ultrasound and magnetic resonance imaging (MRI). Imaging with EM energy near the visible spectrum such as ultra violet (UV) and near infra red (NIR) is commonly achieved with Silicon detectors and traditional lenses. Imaging with shorter wavelengths includes X-Ray (with added modalities such as CT). Imaging with longer wavelengths include infrared, microwave~\cite{charvat2015time}, Thz~\cite{redo2016terahertz}, mm-wave~\cite{appleby2007millimeter, wei2015mtrack}, and RF~\cite{wallace2010analysis}. Overcoming the lack of high quality lenses and availability of dense detector arrays is usually achieved by raster scanning, bare sensor imaging~\cite{napier1983very}, and with compressive sensing~\cite{satat2017lensless}. In this work, our goal is to focus on imaging with longer wavelengths and RF in particular.

\subsection{Imaging with RF}

Imaging in RF spectrum has been widely explored for specialized applications. RADAR is the most known application of RF for sensing applications. Recently new applications emerged.

\textbf{Imaging Through Walls} leverages the fact that RF is not highly absorbed by common construction materials such as wood and drywall. Various works explored different parts of the RF spectrum such as ultrawideband~\cite{yang2005see, ralston2010real, peabody2012through} and WiFi~\cite{adib2013see}. Different applications include capturing the shape of a human~\cite{adib2015capturing, zhao2018through} and human computer interaction applications~\cite{bedri2015rflow}. 


\textbf{Motion Tracking} is a specific instance of RF imaging. Because of the relatively longer wavelength it is usually hard to form images, and specific applications such as motion tracking~\cite{kotaru2015spotfi} are tackled directly. Tracking can be used for monitoring vital signs~\cite{adib2015smart}, or for human computer interaction~\cite{kotaru2017position,pu2013whole,joshi2015wideo}.

\textbf{Synthetic Aperture Radar (SAR)} overcomes the main limitation in forming an image with RF. Because of the long wavelength and size of detectors, a very large aperture is required which in practice is not feasible. SAR overcomes that challenge by forming a synthetic large aperture~\cite{ralston2010interferometric, chen2014inverse, holloway2017savi}.

The systems above provide application-specific description of RF imaging.  Here, we suggest a different perspective for imaging with RF. We expand fundamental concepts in visible light imaging systems like light-field to describe RF imaging systems. Our formulation accounts for wave effects like diffraction and interference which play a dominant role in RF imaging systems due to large wavelengths. Wigner Distribution Functions have been considered to describe wave effects in the context of light field for visible light~\cite{zhang2009wigner}. By expanding fundamental concepts in computational photography with visible light to RF, we expect that many novel RF imaging applications will emerge.

\section{Plenoptic Function}\label{sec:lightField}
Light emitted from a signal source travels through the environment and interacts with it. The plenoptic function~\cite{adelson1991plenoptic} is a useful tool to describe light transport. Similarly, a RF plenoptic function is helpful when thinking about RF systems. For completeness, we provide a brief introduction to the concepts of the plenoptic function in optical frequencies. In general, we can describe the light field as the real-valued high-dimensional function:
\begin{equation}
L(x,y,z, \xi, \psi, t, f)
\end{equation}
where $x,y,z$ are the spatial coordinates, $\xi$ and $\psi$ are the azimuth and elevation angles of propagation respectively, $t$ is time, $f$ represents the optical frequency. $L(x,y,z, \xi, \psi, t, f)$ is a measure of the radiance of a ray traveling along the direction dictated by the angles of propagation, at a point in space dictated by the spatial coordinates, at time $t$, carrying a signal of frequency $f$.

The underlying concepts of the plenoptic function in optical frequencies are useful in the RF spectra as well. In this case it is useful to think about wavefronts as opposed to light rays. For example, the angular components now describe the normal to the local wavefront. Also, unlike visible light sources most of which emit incoherent light, most of the RF sources generate coherent signal. In visible light systems, unless the source is a laser producing coherent light, the emitted light contains waves with all possible phase values resulting in a transmitted signal that has no useful phase information. Similarly, typical visible light sensors can only measure the intensity and radiance of the incident light. So, using a real-valued plenoptic function, that ignores phase, to describe the radiance of light is sufficient for visible light based systems. In contrast, most of the RF systems generate coherent waves that are in phase with each other at the signal source. 

\begin{figure}
  \centering
      \includegraphics[width=0.9\linewidth]{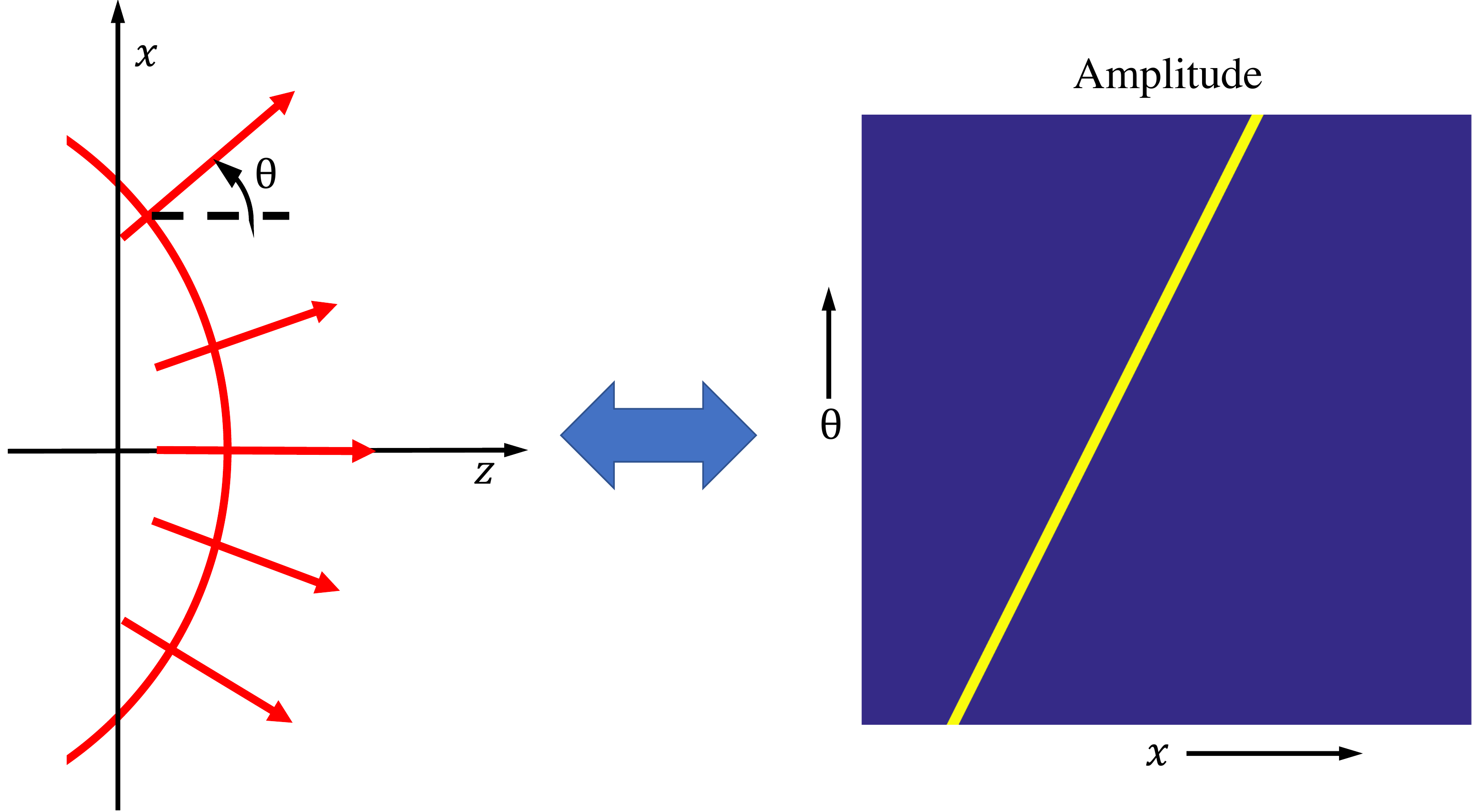}
  \caption{An example illustration of RF plenoptic function. (Left) Illustration of spherical wavefront and rays on a two-dimensional plane. (Right) The amplitude of RF plenoptic function as a function of two of the input variables -- the x coordinate and angle of propagation.}
  \label{fig:spherical}
\end{figure}

The complex-valued RF signal, containing both amplitude and phase, as a function of three-dimensional position, two-dimensional propagation angle, time and frequency is surprisingly amenable to mathematical treatment similar to that of light plenoptic function. We call this the RF plenoptic function $L(x,y,z, \xi, \psi, t, f)$. $L(x,y,z, \xi, \psi, t, f)$ is the complex RF signal at a point $(x,y,z)$, whose normal at the local wavefront at those spatial coordinates is pointing along the direction dictated by the azimuth and elevation angles $\xi$ and $\psi$ respectively, at time $t$, carrying a signal of frequency $f$. Note that unlike light plenoptic function which is real-valued, RF plenoptic function is complex-valued. Fig.~\ref{fig:spherical} shows an example plenoptic function for a simple scenario of a spherical wave propagating in space.

\section{Camera}\label{sec:sensor}
\textbf{Image: }An image captured by a camera is a projection of the plenoptic function. A pinhole camera captures a 2D projection of the plenoptic function $L(x_0,y_0,z_0, \xi, \psi, t_0)$ where $(x_0,y_0,z_0)$ is the position of pinhole, $t_0$ is the time of acquisition and the plenoptic function is integrated over all the frequencies. So, the captured image is just a function of propagation angles $\xi$ and $\psi$. Similarly, traditional and ubiquitous visible light cameras are 2D sensor arrays and generate a 2D projection of the plenoptic function:
\begin{itemize}
\item Each sensor integrates over all angles defined by the camera aperture. 
\item The depth information is lost in the projection of the scene on the sensor. 

\item The camera integrates over time during the exposure. 
\item A Bayer mask is used to sample different wavelengths (usually, these filters are broad such that each pixel integrates over a fraction of the spectrum). 
\end{itemize} 

There are many variants to the traditional camera. For example, a light-field camera captures the angular information (usually at the cost of reduced spatial resolution). A spectrometer provides high spectral resolution (at the cost of spatial information). Ultrafast cameras provides high temporal resolution and so on. 

In general, due to technological limitations only a few parameters of the plenoptic function can be sampled with high resolution. The choice of the important parameters is determined by the application. For example, in traditional cameras the spatial coordinates are sampled with high resolution.

Due to significant difference between optical and radio frequencies and the resulting differences in visible light and RF camera systems, our ability to sample the RF plenoptic function is substantially different. RF camera samples the complex-valued RF plenoptic function generating complex-valued measurements. The projection captured by an RF camera, e.g. a WiFi receiver, can be written as $L(x_k,y_k,z_k, f)$:
 \begin{itemize}
\item RF camera has a sensor array consisting of $K$ sensors and $(x_k,y_k,z_k)$ is the position of $k^{\mathrm{th}}$ sensor.
\item The sensors are typically omni-directional. Each sensor integrates over all the propagation angles. So, unlike visible light cameras, angle dependent information of the plenoptic function is completely lost. 
\item The camera integrates over time. 
\item A RF filter is used to isolate the signal from each of the constituent frequencies $f$. 
\end{itemize}

\textbf{Pixels: }Pixels are the sensing elements of a camera. A photodector is the sensing element of a visible light camera. It integrates the light incident on it from all the directions within its field of view across time and produces a current that generates a single reading or measurement. RF Pixels are antennas with accompanying circuits that perform a similar operation as a photodetector measuring the incident RF signal at the antenna from all the directions across time and generating a reading. The antennas can be arranged in 1D, 2D, or 3D structures. RF pixels have a much larger field of view than photodetectors. For example, antennas used in typical RF systems like WiFi in smartphones are omni-directional, i.e., they have a 360 degree field of view in azimuthal direction.

Unlike visible light cameras where millions of pixels can be manufactured in CCD/CMOS arrays, RF spectrum lacks similar processes to manufacture large arrays of RF pixel sensors. Further, the size of RF antennas are comparable to the wavelength and the associated circuitry for filtering and sampling is complex. As a result it is very challenging to create dense array of RF pixels. So, unlike visible light cameras which are equipped with millions of pixels, typical RF cameras are equipped with 3-4 RF pixels. On the other hand, RF pixels measure the phase of the signal in addition to amplitude and the integration times are much shorter.

\textbf{Lens: }An important aspect of imaging in optical frequencies is the use of a lens. A lens transforms the incident light that is coming from all directions such that light from one focused target position is incident on only one pixel. The lens effectively creates a mapping between spatial coordinates in the scene to pixels on the sensor array. More specifically its a projection  $\Pi(x,y,z) \rightarrow (\bar{x},\bar{y})$. If the object is planar, parallel to the aperture, and focused by the lens then the mapping is one-to-one.

It is impractical to build a high-quality lens in RF spectrum due to large wavelengths of RF waves. So, there is no structure like in visible light cameras that maps the positions of the physical world to the pixels in the camera. Thus, we cannot sample RF plenoptic function as a function of propagation angle and each pixel integrates the RF signal incident on it from all the directions.

In summary, RF and traditional cameras compliment each other in the way they sample the plenoptic function. Traditional cameras provide high sampling rates of the spatial coordinates, with poor spectral and phase information. RF cameras on the other hand provides a sparse measurement of the spatial coordinates, but provide high sampling rate of time (frequency) and phase. Another key difference is the lack of a lens in RF; all the signals from different angular coordinates in the scene contribute to the signal measured by an RF pixel.

\section{Emitter}\label{sec:source}
Many RF imaging systems employ active emitters to gain advantage in terms of improving the signal strength and reducing the interference from the ambient sources. RF sources can be broadly classified into three classes:

\noindent\textbf{Continuous wave based: }These sources emit RF light-field with single frequency. This frequency is called as carrier frequency.  Examples of continuous wave sources include Doppler radars used for identifying the speed of moving cars.

\noindent\textbf{Discrete frequency based: }RF light-field emitted from these sources consists of multiple discrete frequencies. Examples include WiFi systems.

\noindent\textbf{Pulse based: }These sources emit short pulses in radio frequency range of electromagnetic spectrum. Examples include pulse-Doppler radars used for see-through-wall applications. 

\begin{figure}
  \centering
      \includegraphics[width=1\linewidth]{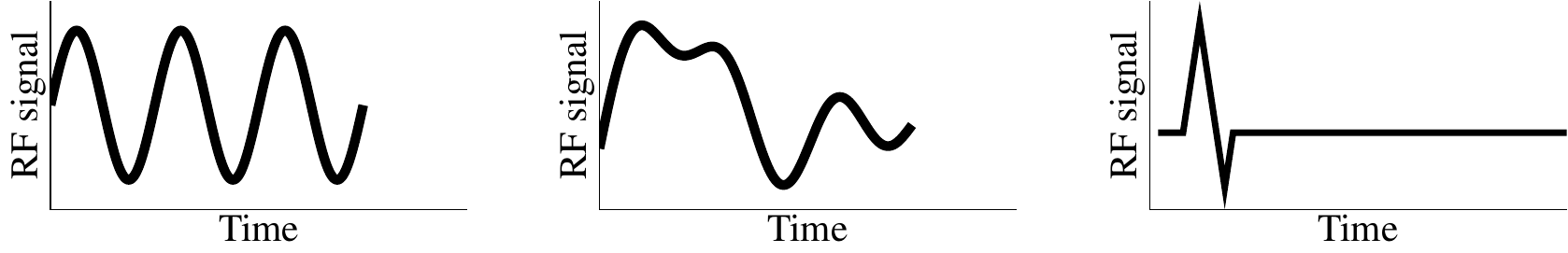}
  \caption{Illustration of signals used in continuous wave RF sources (Left), discrete frequency based sources (Middle) and pulse emitters (Right). Only the real part of the signal is shown.}
  \label{fig:emitter}
\end{figure}

Signal generated by any RF emitter can be mathematically represented as $A\mathrm{e}^{ i2 \pi f_c t } x(t)$. $f_c$ is called as the carrier frequency. The carrier frequency is in the radio frequency portion of electromagnetic spectrum, i.e., between 3~kHz and 300~GHz. $A$ is the complex amplitude of the carrier wave. $x(t)$ is the modulating signal that is modulated over the carrier wave. In continuous wave sources, $x(t)$ is just a constant function. In discrete frequency sources, $x(t)$ is sum of complex sinusoidal signals belonging to a small number of frequencies. In pulse emitters, $x(t)$ is a short, time-limited pulse. Fig.~\ref{fig:emitter} illustrates representative RF signal emitted by these different kinds of sources. Bandwidth of the modulating signal is system-dependent. For example, modulation frequencies used in WiFi are on the order of few 10s of MHz and the bandwidth of the modulating signal used in pulse-Doppler radars is few GHz.

The complex-valued modulating signal $x(t)$ can be decomposed into its constituent Fourier frequency components. In environments which exclusively consist of objects that do not change the frequency of the incident RF light-field, each frequency can be considered separately. In the following sections, each frequency $f$ constituting the emitted signal $A\mathrm{e}^{ i2 \pi f_c t } x(t)$ is considered separately and the corresponding wavelength is denoted by $\lambda$.

\subsection{Relation to active emission in visible light systems}\label{sec:emitLight}
The mathematical representation of signals generated by RF emitters is very similar to that used for describing active illuminators in Time-of-Flight (ToF) cameras. The light generated by these emitters can be represented as $A x(t)$. Since typical light sources, including those used in ToF cameras, are incoherent, the carrier phase term, $\mathrm{e}^{ i2 \pi f_c t }$, is absent. Further, $A$ represents the intensity emitted by the source rather than the amplitude. The modulating signal $x(t)$ is composed of a single modulating frequency, $(1 + \sin(2 \pi f t ) )$~\cite{heide2015doppler}. Note that $x(t)$ modulates amplitude of carrier signal but not the phase. The modulation frequency $f$ is on the order of 10s of MHz for ToF cameras. In pulsed laser ToF systems, $x(t)$ is a pulse with a bandwidth of few hundreds of GHz.

\section{Propagation}\label{sec:free}

Amplitude of RF signal emitted from a point source falls off as $1/r$ as it travels a distance $r$ in free space. Further, the signal undergoes a delay of $r/c$, where $c$ is the speed of light, transforming the signal to $\frac{A}{r} \mathrm{e}^{ i2 \pi f (t - \frac{r}{c})}$. The signal propagates through spherical wavefronts. So, the RF light-field observed at a point $r \mathrm{\textbf{v}}$ with a point source located at the origin, with a unit vector $\mathrm{\textbf{v}}$ , is given by 
\begin{equation}\label{eq:free}
L[x, y, z, \xi, \psi, t , f ] = \frac{A}{r} \mathrm{e}^{ i2 \pi f t} \mathrm{e}^{- i2 \pi \frac{r}{\lambda}} g(\xi, \psi,  \mathrm{\textbf{v}}).
\end{equation}
In Eq.~\ref{eq:free}, $ (x,y,z) = r \mathrm{\textbf{v}}$ and $g(\xi, \psi,  r \mathrm{\textbf{v}})$ is a geometric factor that is 1 only if the direction determined by the angles of interest at the point, $(\xi, \psi)$ is along the unit vector $\mathrm{\textbf{v}}$ joining the source to that point. 

In visible light systems like those discussed in Sec.~\ref{sec:emitLight}, the intensity of the emitted light drops as inverse square of distance of propagation resulting in $\frac{A}{r^2} x(t - \frac{r}{c})$.

\begin{figure}
  \centering
    \includegraphics[width=0.4\linewidth]{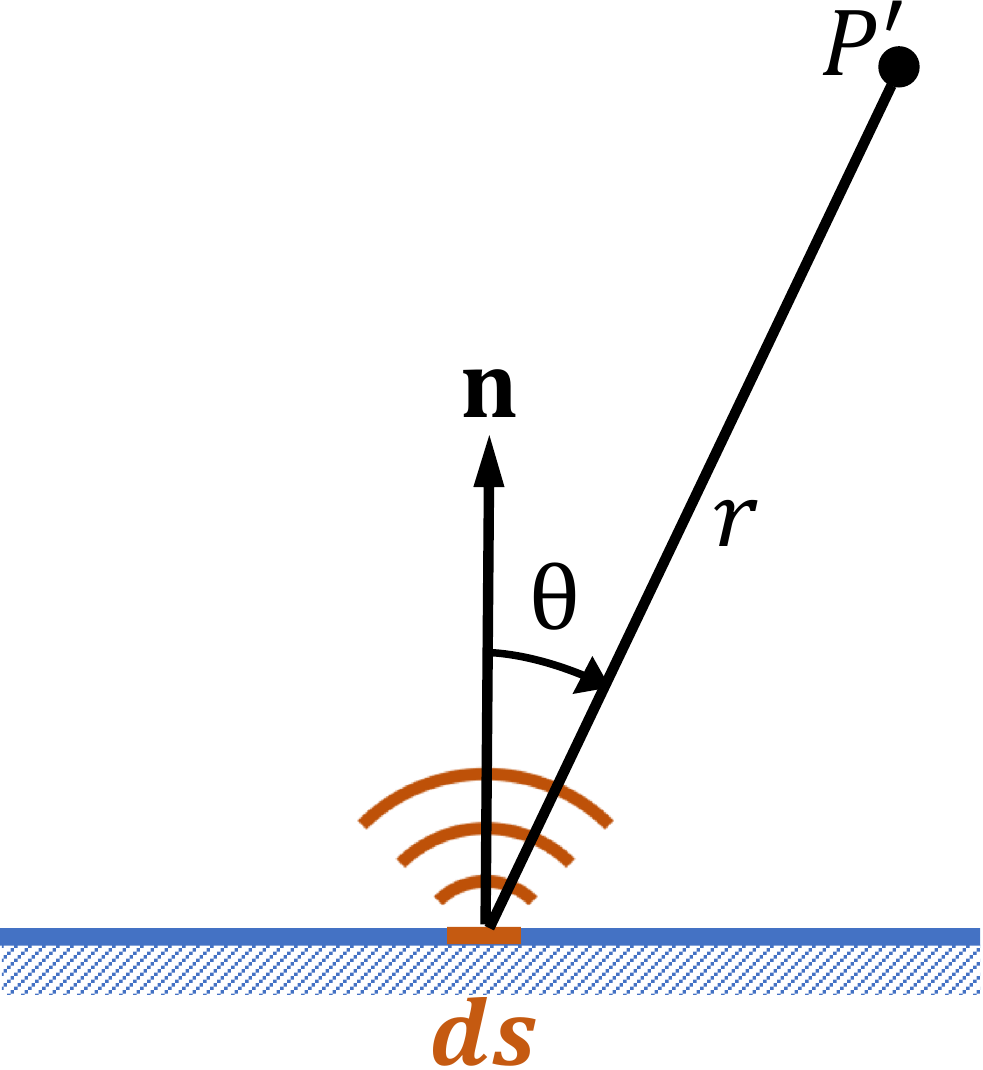}
      \caption{RF signal from a patch is superposition of the signals from wavelet sources.}
      \label{fig:patch}
\end{figure}

\section{Transformation of RF light field}\label{sec:transform}
We will now represent the transformation of RF light-field when it reaches an object or obstruction. For example, consider the interaction of a spatially-uniform but time-dependent RF signal, $\mathrm{e}^{i2\pi f t}$, as it interacts with a thin patch of an object shown in Fig.~\ref{fig:patch}. The RF signal received at a point $P'$ is given by Rayleigh-Sommerfeld formula:
\begin{equation}\label{eq:rayleigh}
u(P') =  \int_S \mathrm{e}^{i2\pi f t} \alpha \frac{\mathrm{e}^{-i 2 \pi r / \lambda}}{r}  \cos(\theta)    ds,
\end{equation}
where $S$ is the surface of the patch, $r$ is the distance between an infinitesimal patch element $ds$ and $P'$, $\theta$ is the angle between the outward normal of the patch and the vector joining the patch element $ds$ to $P'$, and $\alpha$ is a weighting factor which depends on the properties of the material constituting the patch and the wavelength of light. The infinitesimal patch elements are called as wavelet sources.

\begin{figure}
  \centering
    \includegraphics[width=0.8\linewidth]{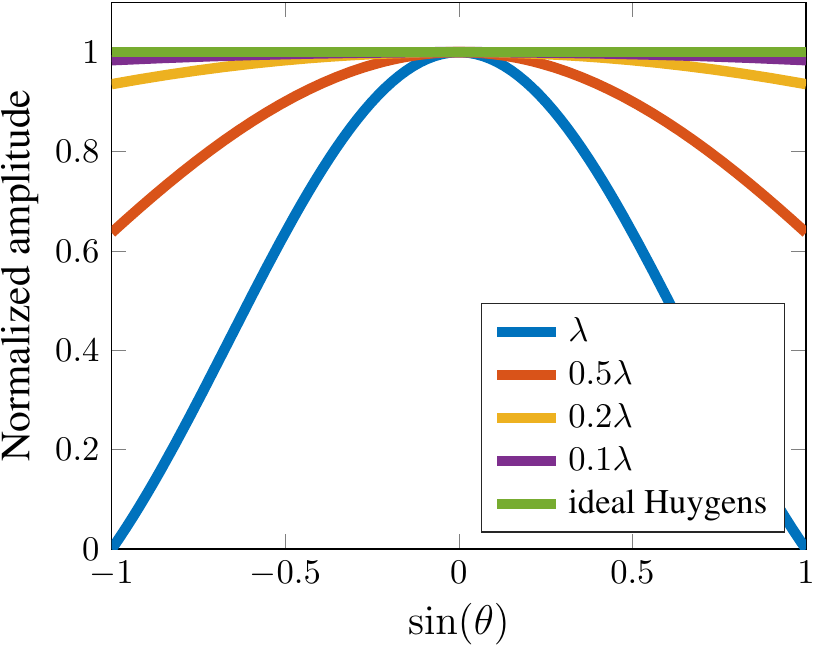}
      \caption{Amplitude of signal from the patch versus the angle of propagation as a function of the patch width.}
      \label{fig:pointApprox}
\end{figure}

If the width of the patch is much smaller than the the distance between the patch and the receiver point $P'$, then the signal $u(P')$ from Eq.~\ref{eq:rayleigh} can be approximated by:
\begin{equation}\label{eq:simple_rayleigh}
u(P') = \mathrm{e}^{i2\pi f t} \alpha  {\mathrm{e}^{-i 2 \pi R / \lambda}} \frac{\cos(\theta)}{R} \int_S {\mathrm{e}^{i 2 \pi s \sin\theta / \lambda}} ds,
\end{equation}
where $R$ is the distance between the midpoint of the patch and the receiver point $P'$, $s$ is the distance from that midpoint to an infinitesimal patch element. The value of the integrand as a function of propagation angle $\theta$ for different sizes of the patch is provided in Fig.~\ref{fig:pointApprox}. The value is normalized by the integrand's value for $\theta=0$. One can clearly observe from Fig.~\ref{fig:pointApprox} that for a larger patch size (in relation to the wavelength), the signal obtained from a thin patch deviates more from that of an ideal wavelet source. If the width of the patch is less than one-tenth of the wavelength, one can approximate the patch by an ideal wavelet source. 

Evaluating Eq.~\ref{eq:simple_rayleigh} for such patches with size smaller than one-tenth of the wavelength results in 
\begin{equation}\label{eq:patch_transform}
u(P') = \mathrm{e}^{i2\pi f t} \alpha  {\mathrm{e}^{-i 2 \pi R / \lambda}} \frac{\cos(\theta)}{R} S,
\end{equation}
where $S$ is the size of the patch. By comparing Eq.~\ref{eq:patch_transform} and Eq.~\ref{eq:free}, one can interpret that each such patch multiplies the incident RF signal by propagation-angle-dependent term $\alpha S \cos(\theta)$, and then emits the resultant signal to be propagated in free space to reach $P'$.

\begin{figure*}
\begin{subfigure}{.31\textwidth}
  \centering
    \includegraphics[width=0.9\linewidth]{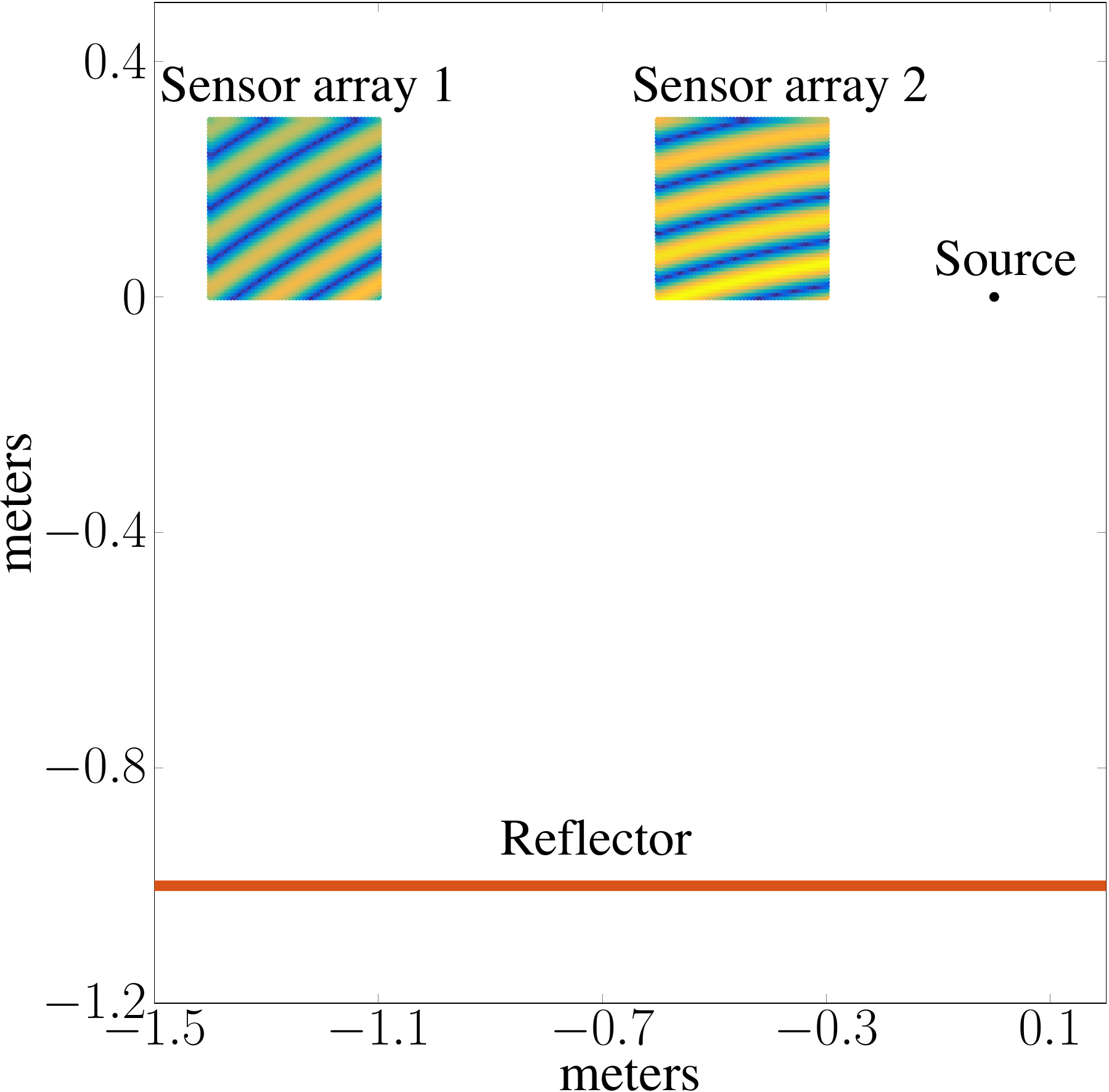}
        \caption{Reflector}
        \label{fig:ref}
\end{subfigure}
\hfill
\begin{subfigure}{.31\textwidth}
  \centering
    \includegraphics[width=0.9\linewidth]{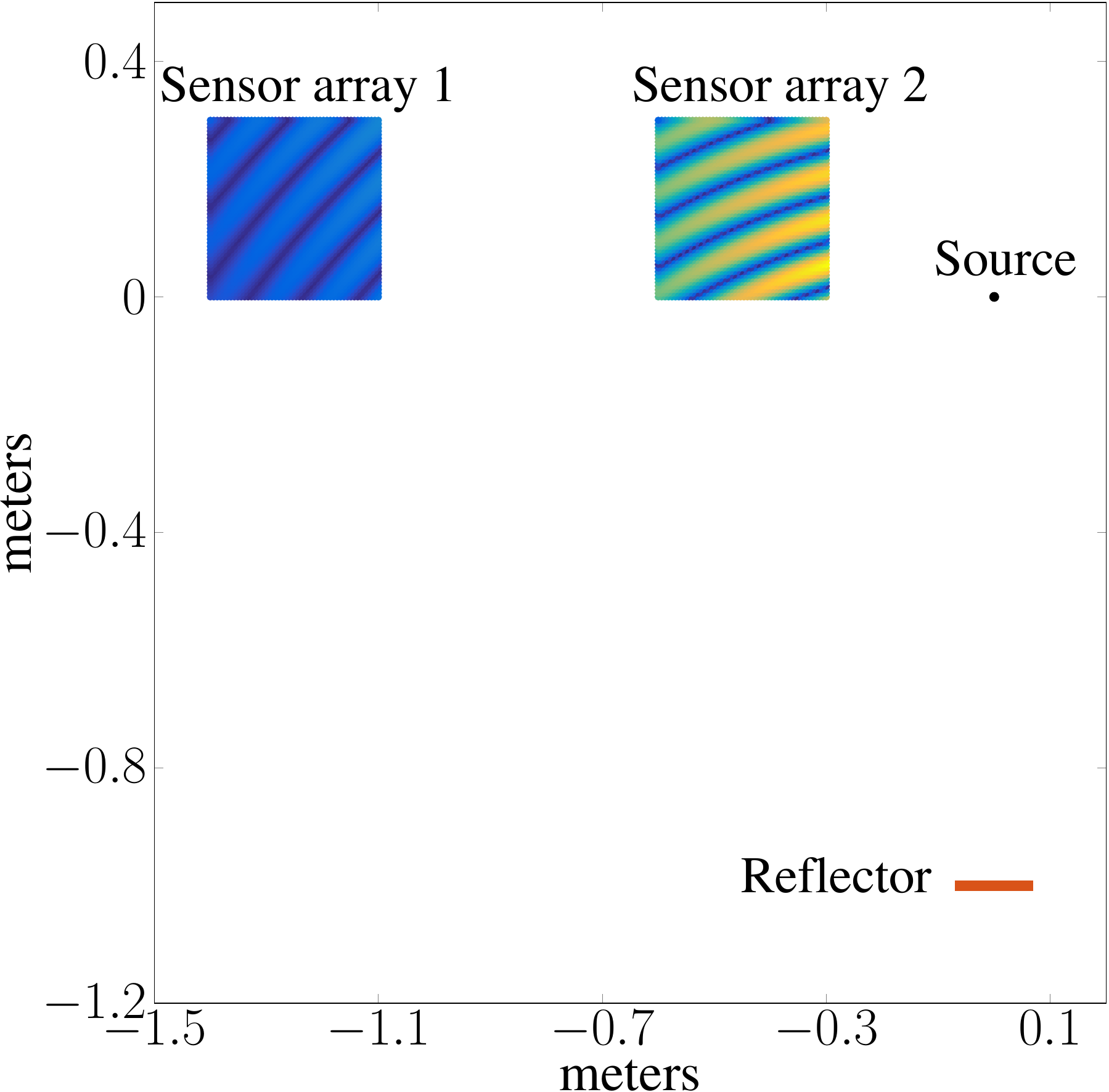}
        \caption{Reflective Diffraction}
        \label{fig:refDiff}
\end{subfigure}
\hfill
\begin{subfigure}{.31\textwidth}
  \centering
    \includegraphics[width=0.9\linewidth]{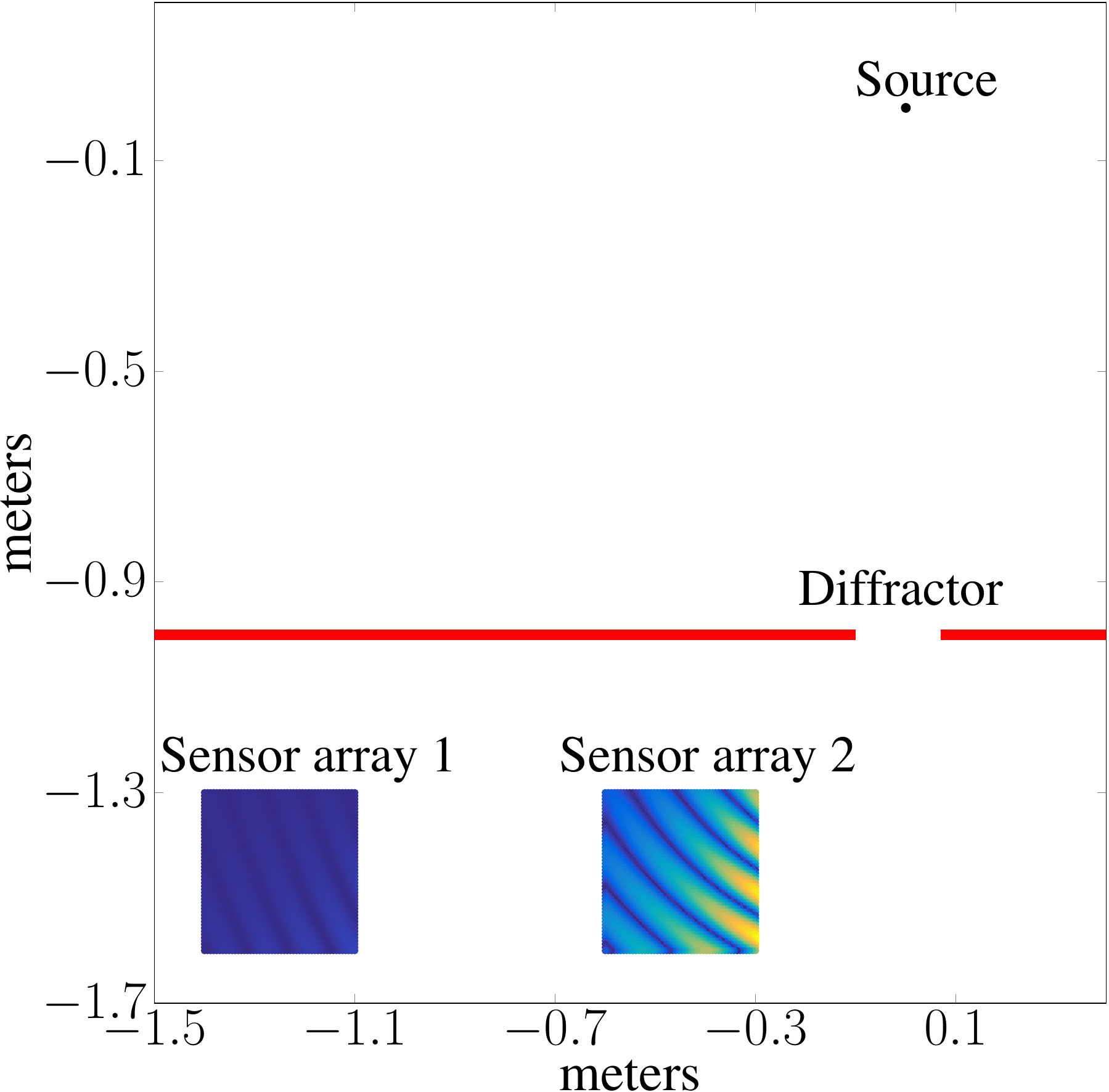}
        \caption{Transmissive Diffraction}
        \label{fig:Diff}
\end{subfigure}
\caption{RF images rendered using superposition of signals from discretized patches approximated as wavelet sources.}
    \label{fig:superpos}
\end{figure*}

Applying this interpretation automatically accounts for different wave effects like reflection, refraction and diffraction. For example, Fig.~\ref{fig:ref} shows the RF signal contributed at an array of points due to a large reflector of size 4~m present in the scene. The frequency of the signal used is 2.4~GHz corresponding to a wavelength of 12.5~cm. The object has been divided into patches of size 1~cm. One can observe that the wavefront arrives at different angles at different arrays of points which is intuitive because signal reflected from large objects follows Newtonian laws of reflection. If the object size is comparable to wavelength of light as in Fig.~\ref{fig:refDiff}, reflective diffraction is observed. Fig.~\ref{fig:Diff} considers a scene in which the source and the array of points are separated by a large object which has a small opening. The opening can be treated as a diffractor object and Eq.~\ref{eq:patch_transform} can be used for calculating signal passing through patches of this diffractor object. In this section, for simplicity, we considered only two-dimensional scenes. However, the concepts can be trivially extended to three-dimensional scenes.

RF signal incident from any angle at a patch reflects in all directions away from the interior of the patch according to Eq.~\ref{eq:patch_transform}. The ratio of reflected signal amplitude compared to incident signal amplitude is given by 
\begin{equation}\label{eq:rf_brdf}
 \alpha S \cos(\theta),
\end{equation}
which provides the reflectance function for RF. This is similar to the concept of BRDF in visible light. Note that this reflectance function for RF is a function of the outgoing direction alone and not the incoming direction. However, we note that this formulation for reflectance function is a result of simplistic model that does not account for sub-surface scattering. More complicated models that account for scattering and procedures for experimentally measuring them of real objects is beyond the scope of this paper and is part of future work.

Eq.~\ref{eq:patch_transform} shows the signal emitted from a patch for an incoming signal arriving from one direction. In general, RF light-field arriving at a patch has signal along multiple directions. The overall emitted RF light-field from the patch can be written as
\begin{equation}
L(x,y,z, \xi, \psi, t, f) = \rho(\xi, \psi) \int\displaylimits_{\xi', \psi'} L(x,y,z,\xi',\psi',t,f)  d\xi' d\psi',
\end{equation}
where $L(x,y,z,\xi',\psi',t,f)$ is the RF light-field incident along direction dictated by azimuth and elevation  angles $\xi'$ and $\psi'$respectively.  $\rho(\xi, \psi)$ is the reflectance function of the patch, Eq.~\ref{eq:rf_brdf}, evaluated along the direction dictated by angles $\xi$ and $\psi$.

\section{Receiver}\label{sec:receiver}
As discussed in Sec.~\ref{sec:sensor}, each antenna sensor of a RF camera integrates RF light-field over all angles at the RF camera. However, an RF camera does not simply sample the resultant signal at different time instants. An RF camera multiplies the signal captured by antennas with a reference sinusoidal signal before integrating the signal over exposure time. This process is called demodulation. 

Let $L[x_k, y_k, z_k, \xi, \psi, t, f]$ be the RF light-field incident at an antenna located at $(x_k, y_k, z_k)$. The antenna of RF camera integrates RF light-field along all angles and then performs demodulation to obtain the image
\begin{equation}\label{eq:receiver}
I(x_k, y_k, z_k,f) = \int_T \Bigg( \int_{\theta,\psi} L[x_k, y_k, z_k, \xi, \psi, t, f]  \Bigg) \mathrm{e}^{-i2\pi f t - \phi},
\end{equation}
where $T$ is the integration time, and $\xi$ and $\psi$ are propagation angles at RF camera. The exposure time varies from a fraction of a microsecond in pulse-based radars to few microseconds in WiFi-based RF cameras. 


Notice that the sinusoidal signal used for demodulation $\mathrm{e}^{-i2\pi f t - \phi}$ has the same frequency as the signal used at the emitter. However, the phase of the sinusoidal signal can differ from that of the phase used in transmitted signal. In RADAR systems, both the emitter and receiver are on the same device  and share the same oscillator that generates the sinusoidal signal; so, in radars $\phi$ is 0. But in systems like WiFi, the emitter and receiver are separate devices and the relative phase $\phi$ between the sinusoidal signal used at the emitter and receiver cannot be controlled or known.

\subsection{Relation to receiver in visible light cameras}
In traditional cameras the light field is integrated during the exposure time. ToF cameras like Kinect perform a simple demodulation operation by multiplying the received signal with ${\sin(2\pi ft + \phi)}$ and then passing the resultant signal through a low pass filter by integrating the signal over exposure time $T$. $T$ is typically few milliseconds. The relative phase $\phi$ is known and accurately controlled as both the emitter and receiver are on the same device and share the oscillator generating the sinusoidal signal.

\begin{figure*}
\begin{subfigure}{.24\textwidth}
  \centering
    \includegraphics[width=1\linewidth]{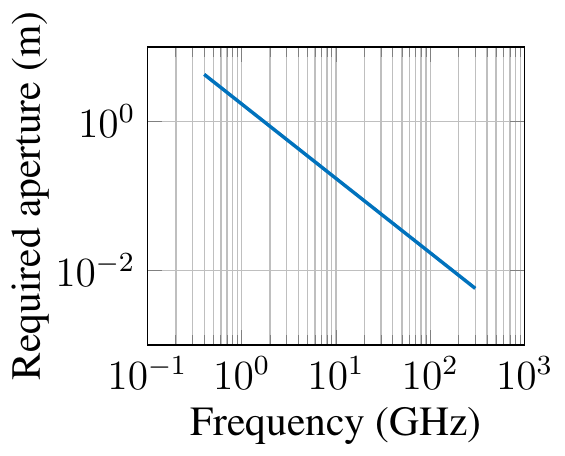}
      \caption{Length of array required for achieving an angular resolution of 10 degrees as a function of frequency.}
      \label{fig:arrayRes}
\end{subfigure}
\hfill
\begin{subfigure}{.24\textwidth}
  \centering
    \includegraphics[width=1\linewidth]{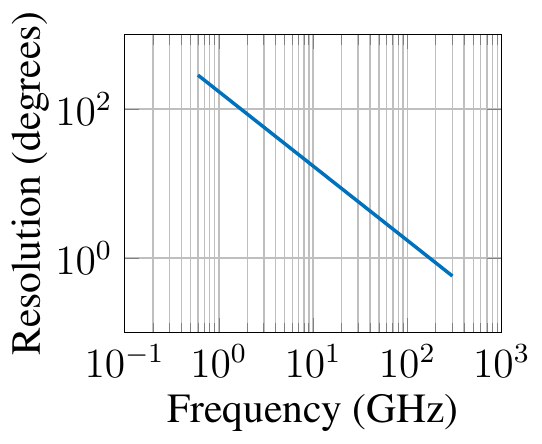}
      \caption{Resolution of 10 cm long array as a function of frequency.}
      \label{fig:fixedArrayRes}
\end{subfigure}
\hfill
\begin{subfigure}{.24\textwidth}
  \centering
    \includegraphics[width=1\linewidth]{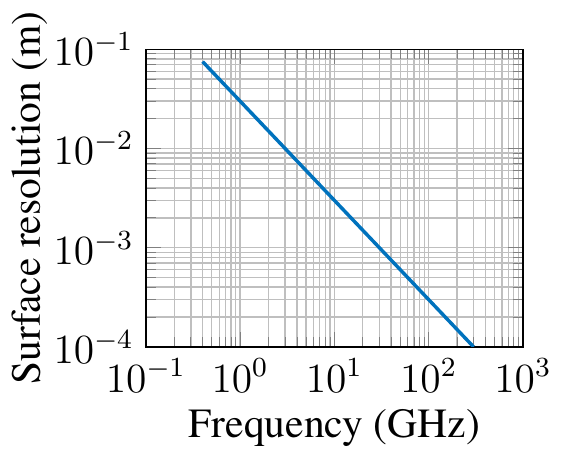}
      \caption{Resolution limit in imaging surface irregularities as a function of frequency.}
      \label{fig:surfRes}
\end{subfigure}
\hfill
\begin{subfigure}{.24\textwidth}
  \centering
    \includegraphics[width=1\linewidth]{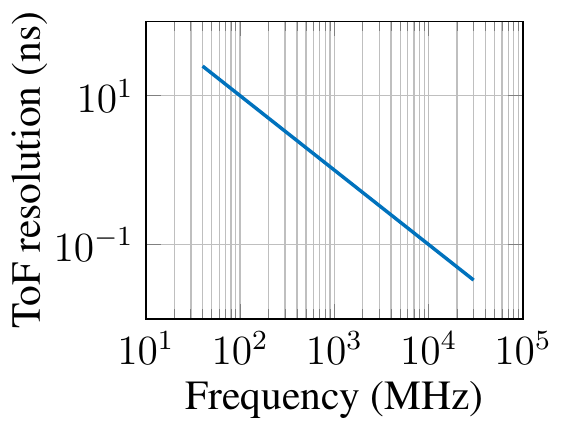}
      \caption{Time-of-Flight Resolution as a function of bandwidth.}
      \label{fig:tofRes}
\end{subfigure}
\caption{Analysis of resolution}
\end{figure*}

\section{Imaging Resolution}\label{sec:resolution}

In this section, we analyze the imaging resolution provided by a RF camera. The images produced at different antennas and at different frequencies is given by Eq.~\ref{eq:receiver}. The information of RF light-field as a function of propagation angles is embedded in that image information. But, it is not readily accessible, like in a traditional visible camera where a lens maps the signal from different scene points to different pixels in an image. So, computational algorithms like synthetic aperture radar are used to recover a more traditional representation of image, i.e., reflected intensity as a function of Cartesian coordinates of the imaged scene. We now analyze the resolution of such representation in simple two-dimensional scenes. However, the analysis can be extended to three-dimensional scenes.

\noindent\textbf{Aperture: } Consider a dense array of antennas in a linear array where adjacent antennas are separated by no more than half the wavelength of light. The length of the antenna array is called an aperture. As shown in~\cite{bouman2016computational}, the resolution with which signals from different angles can be resolved in a dense scene, i.e., the RF signal has significant value for every angle at the camera, is given by 
\begin{equation}\label{eq:array_res}
\mathrm{resolution} = \frac{\lambda}{\mathrm{aperture}}.
\end{equation}
The length of the array required for angular resolution of 10~degrees for different frequencies in RF spectrum in plotted in Fig.~\ref{fig:arrayRes}. Similarly, the resolution offered by a 10~cm array in different frequencies is presented in Fig.~\ref{fig:fixedArrayRes}. Note that in the algorithms used to achieve the resolution dictated by Eq.~\ref{eq:array_res}, it is assumed that the range to the object is very large when compared to the aperture. 

\noindent\textbf{Wavelength: }It is impossible to recover arbitrarily fine features like irregularities on a surface just by increasing the length of the antenna array. As shown in Fig.~\ref{fig:pointApprox}, if a patch of size one-tenth of wavelength or smaller is composed of uniform material, then the patch can be approximated by a wavelet point source. So, once the resolution dictated by Eq.~\ref{eq:array_res} reaches this limit, then increasing the array length does not enable to image finer surface irregularities. This is because there is no difference between such a surface and a point source. This limit in the resolution as a function of frequencies in radio spectrum is shown in Fig.~\ref{fig:surfRes}. We note that this limit is applicable only if the surface is composed of material with same RF properties. 


\noindent\textbf{Bandwidth: }The signal from different elements in a scene can also be resolved using Time-of-Flight. So, even if the signal from two elements in a scene have close-by angles with respect to the RF camera, they can be resolved in the time dimension. This is possible because the RF image is obtained at different frequencies. The time resolution as a function of bandwidth, i.e., the range of frequencies used in the imaging system, is shown in Fig.~\ref{fig:tofRes}.

\noindent\textbf{Object size: }Small objects act as reflective diffractors and reflect in a diffuse manner. However, the amplitude of reflected light is small because the size of the object is small. Large reflectors reflect in a specular manner. Such that the intensity of light from the object is focused only in one direction and the signal is very small in all other directions. We analyze this interplay here by calculating amplitude of light reflected from an object. We consider a scenario where the emitter and the receiver are at the same location, like a RADAR. The object is placed parallel to the antenna array. The \textit{angle of the object} is defined as the angle of the vector joining the midpoint of the object and the midpoint of the array with respect to the normal of the antenna array. The frequency of the RF signal is set at 2.4~GHz. The object plane is assumed to be 4~m away from the RF camera plane.

\begin{figure}[h]
  \centering
    \includegraphics[width=0.8\linewidth]{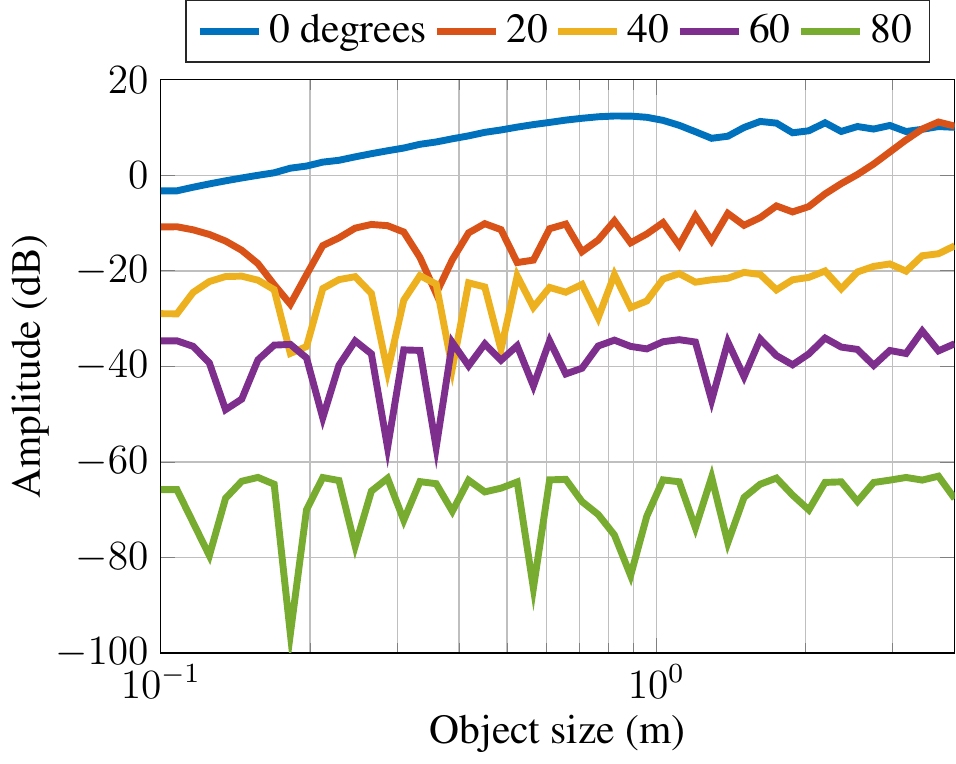}
      \caption{Amplitude of the reflected signal as a function of the object size and angle of the object.}
      \label{fig:sameAngle}
\end{figure}

Fig.~\ref{fig:sameAngle} plots the amplitude of the RF signal obtained at the midpoint of antenna array. From the figure, we observe that when the \textit{angle of the object} is small, increasing the size of the object results in higher amplitude. However, if the \textit{angle of the object} is large, then the effects of increased size and increased reflector specularity tend to cancel out as the object size increases.


\section{Discussion}

We presented the fundamentals towards exploiting computational photography concepts in RF cameras. The current formulation does not account for several interactions. We plan to explore the following limitations in a future work:
\begin{itemize}
    \item \textbf{Scattering:} In the interaction with reflectors, we assumed a simple  model for reflectance function that does not account for any sub-surface scattering. Another example is the lack of scattering in the propagation model.
    
    \item \textbf{Polarization:} We assumed the RF light-field is a scalar and ignored polarization effects. In practice, different materials and interactions may change the polarization state of the RF light-field.
    
    \item \textbf{Non linear interactions:} Certain reflectors may change the frequency of the emitted RF light-field. Here, we assumed that the frequency parameter does not change throughout the interactions.
\end{itemize}

In summary, we proposed the RF light-field that expands the concept of the Plenoptic function to RF. We developed several transformations including emission, propagation, interaction, and detection in RF as well as a rendering procedure for RF. Several fundamental imaging dynamics were demonstrated such as imaging resolution as a function of aperture size and system bandwidth.

\bibliographystyle{IEEEtran}
\bibliography{egbib}

\begin{thebibliography}{10}
\providecommand{\url}[1]{#1}
\csname url@samestyle\endcsname
\providecommand{\newblock}{\relax}
\providecommand{\bibinfo}[2]{#2}
\providecommand{\BIBentrySTDinterwordspacing}{\spaceskip=0pt\relax}
\providecommand{\BIBentryALTinterwordstretchfactor}{4}
\providecommand{\BIBentryALTinterwordspacing}{\spaceskip=\fontdimen2\font plus
\BIBentryALTinterwordstretchfactor\fontdimen3\font minus
  \fontdimen4\font\relax}
\providecommand{\BIBforeignlanguage}[2]{{%
\expandafter\ifx\csname l@#1\endcsname\relax
\typeout{** WARNING: IEEEtran.bst: No hyphenation pattern has been}%
\typeout{** loaded for the language `#1'. Using the pattern for}%
\typeout{** the default language instead.}%
\else
\language=\csname l@#1\endcsname
\fi
#2}}
\providecommand{\BIBdecl}{\relax}
\BIBdecl

\bibitem{charvat2015time}
G.~Charvat, A.~Temme, M.~Feigin, and R.~Raskar, ``Time-of-flight microwave
  camera,'' \emph{Nature Scientific reports}, vol.~5, 2015.

\bibitem{adib2015capturing}
F.~Adib, C.-Y. Hsu, H.~Mao, D.~Katabi, and F.~Durand, ``Capturing the human
  figure through a wall,'' \emph{ACM Transactions on Graphics (TOG)}, vol.~34,
  2015.

\bibitem{redo2016terahertz}
A.~Redo-Sanchez, B.~Heshmat, A.~Aghasi, S.~Naqvi, M.~Zhang, J.~Romberg, and
  R.~Raskar, ``Terahertz time-gated spectral imaging for content extraction
  through layered structures,'' \emph{Nature communications}, vol.~7, p. 12665,
  2016.

\bibitem{appleby2007millimeter}
R.~Appleby and R.~N. Anderton, ``Millimeter-wave and submillimeter-wave imaging
  for security and surveillance,'' \emph{Proceedings of the IEEE}, vol.~95,
  2007.

\bibitem{wei2015mtrack}
T.~Wei and X.~Zhang, ``mtrack: High-precision passive tracking using millimeter
  wave radios,'' in \emph{Proceedings of the 21st Annual International
  Conference on Mobile Computing and Networking}, 2015.

\bibitem{wallace2010analysis}
H.~B. Wallace, ``Analysis of rf imaging applications at frequencies over 100
  ghz,'' \emph{Applied optics}, vol.~49, no.~19, pp. E38--E47, 2010.

\bibitem{napier1983very}
P.~J. Napier, A.~R. Thompson, and R.~D. Ekers, ``The very large array: Design
  and performance of a modern synthesis radio telescope,'' \emph{Proceedings of
  the IEEE}, vol.~71, 1983.

\bibitem{satat2017lensless}
G.~Satat, M.~Tancik, and R.~Raskar, ``Lensless imaging with compressive
  ultrafast sensing,'' \emph{IEEE Transactions on Computational Imaging},
  vol.~3, no.~3, pp. 398--407, 2017.

\bibitem{yang2005see}
Y.~Yang and A.~E. Fathy, ``See-through-wall imaging using ultra wideband
  short-pulse radar system,'' in \emph{IEEE Antennas and Propagation Society
  International Symposium}, vol.~3, 2005.

\bibitem{ralston2010real}
T.~S. Ralston, G.~L. Charvat, and J.~E. Peabody, ``Real-time through-wall
  imaging using an ultrawideband multiple-input multiple-output (mimo) phased
  array radar system,'' in \emph{IEEE International Symposium on Phased Array
  Systems and Technology (ARRAY)}, 2010.

\bibitem{peabody2012through}
J.~E. Peabody, G.~L. Charvat, J.~Goodwin, and M.~Tobias, ``Through-wall imaging
  radar,'' \emph{Lincoln Lab. J}, vol.~19, 2012.

\bibitem{adib2013see}
F.~Adib and D.~Katabi, \emph{See through walls with WiFi!}\hskip 1em plus 0.5em
  minus 0.4em\relax ACM, 2013, vol.~43, no.~4.

\bibitem{zhao2018through}
M.~Zhao, T.~Li, M.~Abu~Alsheikh, Y.~Tian, H.~Zhao, A.~Torralba, and D.~Katabi,
  ``Through-wall human pose estimation using radio signals,'' in
  \emph{Proceedings of the IEEE Conference on Computer Vision and Pattern
  Recognition}, 2018, pp. 7356--7365.

\bibitem{bedri2015rflow}
H.~Bedri, O.~Gupta, A.~Temme, M.~Feigin, G.~Charvat, and R.~Raskar, ``Rflow:
  User interaction beyond walls,'' in \emph{Adjunct Proceedings of the 28th
  Annual ACM Symposium on User Interface Software \& Technology}, 2015.

\bibitem{kotaru2015spotfi}
M.~Kotaru, K.~Joshi, D.~Bharadia, and S.~Katti, ``Spotfi: Decimeter level
  localization using wifi,'' in \emph{ACM SIGCOMM Computer Communication
  Review}, vol.~45, no.~4.\hskip 1em plus 0.5em minus 0.4em\relax ACM, 2015,
  pp. 269--282.

\bibitem{adib2015smart}
F.~Adib, H.~Mao, Z.~Kabelac, D.~Katabi, and R.~C. Miller, ``Smart homes that
  monitor breathing and heart rate,'' in \emph{Proceedings of the 33rd Annual
  ACM Conference on Human Factors in Computing Systems}, 2015.

\bibitem{kotaru2017position}
M.~Kotaru and S.~Katti, ``Position tracking for virtual reality using commodity
  wifi,'' in \emph{Computer Vision and Pattern Recognition (CVPR), 2017 IEEE
  Conference on}.\hskip 1em plus 0.5em minus 0.4em\relax IEEE, 2017, pp.
  2671--2681.

\bibitem{pu2013whole}
Q.~Pu, S.~Gupta, S.~Gollakota, and S.~Patel, ``Whole-home gesture recognition
  using wireless signals,'' in \emph{Proceedings of the 19th annual
  international conference on Mobile computing \& networking}.\hskip 1em plus
  0.5em minus 0.4em\relax ACM, 2013, pp. 27--38.

\bibitem{joshi2015wideo}
K.~Joshi, D.~Bharadia, M.~Kotaru, and S.~Katti, ``Wideo: fine-grained
  device-free motion tracing using rf backscatter,'' in \emph{Proceedings of
  the 12th USENIX Conference on Networked Systems Design and
  Implementation}.\hskip 1em plus 0.5em minus 0.4em\relax USENIX Association,
  2015, pp. 189--204.

\bibitem{ralston2010interferometric}
T.~S. Ralston, G.~L. Charvat, S.~G. Adie, B.~J. Davis, P.~S. Carney, and S.~A.
  Boppart, ``Interferometric synthetic aperture microscopy: Microscopic laser
  radar,'' \emph{Optics and Photonics News}, vol.~21, 2010.

\bibitem{chen2014inverse}
V.~Chen and M.~Martorella, ``Inverse synthetic aperture radar,'' \emph{Scitech
  Publishing 2014, and references therein}, 2014.

\bibitem{holloway2017savi}
J.~Holloway, Y.~Wu, M.~K. Sharma, O.~Cossairt, and A.~Veeraraghavan, ``Savi:
  Synthetic apertures for long-range, subdiffraction-limited visible imaging
  using fourier ptychography,'' \emph{Science advances}, vol.~3, no.~4, p.
  e1602564, 2017.

\bibitem{zhang2009wigner}
Z.~Zhang and M.~Levoy, ``Wigner distributions and how they relate to the light
  field,'' in \emph{Computational Photography (ICCP), 2009 IEEE International
  Conference on}.\hskip 1em plus 0.5em minus 0.4em\relax IEEE, 2009, pp. 1--10.

\bibitem{adelson1991plenoptic}
E.~H. Adelson, J.~R. Bergen \emph{et~al.}, ``The plenoptic function and the
  elements of early vision,'' 1991.

\bibitem{heide2015doppler}
F.~Heide, W.~Heidrich, M.~Hullin, and G.~Wetzstein, ``Doppler time-of-flight
  imaging,'' \emph{ACM Transactions on Graphics (ToG)}, vol.~34, no.~4, p.~36,
  2015.

\bibitem{bouman2016computational}
K.~L. Bouman, M.~D. Johnson, D.~Zoran, V.~L. Fish, S.~S. Doeleman, and W.~T.
  Freeman, ``Computational imaging for vlbi image reconstruction,'' in
  \emph{Proceedings of the IEEE Conference on Computer Vision and Pattern
  Recognition}, 2016, pp. 913--922.

\end{thebibliography}

\end{document}